\documentclass{ws-procs9x6}

\begin{document}

\title{DIJET RATES WITH SYMMETRIC $E_T$ CUTS}

\author{A.~BANFI}

\address{Dipartimento di Fisica G. Occhialini \\
Universit\`a degli Studi di Milano-Bicocca \\
Piazza della Scienza, 3, 20126 Milano, Italy}

\address{University of Cambridge, Cavendish Laboratory \\ 
Madingley Road, CB3 0HE Cambridge, UK\\ 
E-mail: abanfi@hep.phy.cam.ac.uk}

\address{DAMTP, Centre for Mathematical Sciences \\
Wilberforce Road, CB3 0WA Cambridge, UK}

\maketitle

\abstracts{We discuss the physics underlying an all-order resummation
  of logarithmic enhanced contributions to dijet cross sections, and
  present preliminary results for the distribution in the dijet
  transverse energy difference in DIS.}

\section{Introduction}
\label{sec:intro}
Clustering hadrons into jets is a very useful tool to make QCD
predictions. Measuring jet cross section removes the theoretical
uncertainty due to fragmentation functions, and makes it possible to
directly compare data with perturbative (PT) QCD predictions, whose
degrees of freedom are partons and not hadrons. This comparison works
extremely well when considering high transverse energy ($E_T$) jets.
In this kinematical situation one is expected to probe quarks and
gluons at small distances. This joint theoretical and experimental
effort has led to the measurement of the QCD coupling $\alpha_s$ from
jet inclusive $E_T$ spectra and to better constrain the gluon
density in the proton\cite{ZEUS-multijet}.

There are actual limitations in the use of perturbation theory.
First of all any perturbative series diverges.  This is related to the
fact that the observed degrees of freedom are hadrons and not partons,
and results in an ambiguity in PT predictions that fortunately is
suppressed by inverse powers of the hard scale of the process. The
second limitation has to do with the fact that the coefficients in the
PT expansion can be logarithmically enhanced due to incomplete
real-virtual cancellations in the infrared (IR). This has the
consequence that in particular phase space regions one observes a
breakdown of the PT expansion, which can be cured only by performing
an all-order resummation of such logarithms.

It is therefore mandatory, in order to have predictions valid in the
whole of the phase space, to complement any fixed order calculation
with the all-order resummation of logarithmic enhanced contributions,
and, if possible, to remove the ambiguity in the PT expansion with a
power-suppressed non-perturbative (NP) correction, which, unless it
can be computed from the lattice, has to be taken as a
phenomenological input\cite{DSreview}.

\section{The observable}
\label{sec:observable}
\begin{figure}[htbp]
  \centering
  \centerline{\epsfxsize=.6\textwidth\epsfbox{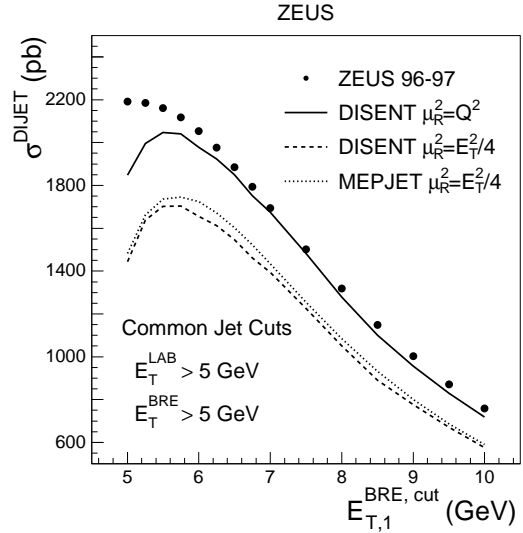}}   
  \caption{The dijet rate $\sigma(\Delta)$ as a function of $E_{T1}$ in DIS. 
    The dots are the data, while the lines are NLO predictions.}
    \label{fig:asym-ZEUS}
\end{figure}

The observable we consider is the dijet rate, i.e. the fraction of
events with at least two jets. To select hard jets we put a cut on the
transverse energy of the two highest $E_T$ jets, requiring
$E_{T1}>E_{T2}>E_m$. This particular choice is referred to as
``symmetric $E_T$ cuts''.  It was noted by Klasen and Kramer\cite{KK}
that symmetric $E_T$ cuts produce IR instabilities in next-to-leading
order (NLO) QCD predictions. It was later Frixione and
Ridolfi\cite{FR} who proposed to perform an asymmetry study by
considering
\begin{equation}
  \label{eq:asym}
  \sigma(\Delta)\equiv\sigma(E_{t1}>E_m+\Delta; E_{t2}>E_m)\>.
\end{equation}
They computed $\sigma(\Delta)$ in photoproduction at NLO, and obtained
that while $\sigma(0)$, the total dijet rate with symmetric $E_T$
cuts, is finite, the slope of the curve $\sigma'(\Delta)\equiv
d\sigma/d\Delta$ diverges for $\Delta=0$.  
This behaviour is not
present at all in the data, as one can see from the asymmetry study
performed by ZEUS in DIS\cite{ZEUS}, and reported in
fig.~\ref{fig:asym-ZEUS}. There the data is plotted against NLO QCD
predictions obtained from the numerical program DISENT\cite{DISENT}.
Again, the slope of the NLO curve diverges for $\Delta=0$, while the
data decreases smoothly with increasing $\Delta$. This can be easily
understood by noting that $\sigma'(\Delta)$ is just (minus) the
differential distribution of the highest $E_T$ jet:
\begin{equation}
  \label{eq:slope}
\sigma'(\Delta) = -\left.\frac{d\sigma}{dE_{T1}}\right|_{E_{T1}=E_m+\Delta}\>.
\end{equation}
Since the latter quantity is a physical cross section, the slope
$\sigma'(\Delta)$ has to be negative. This corresponds to what is seen
in the data, and also implies that any turnover in the NLO curves is
unphysical.

\begin{figure}[htbp]
  \centering
    \centerline{\epsfxsize=.6\textwidth\epsfbox{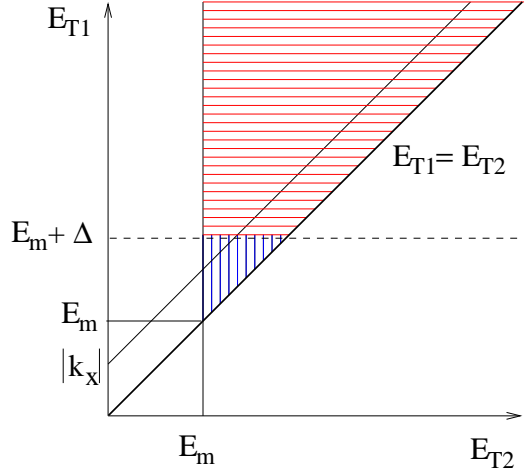}}
    \caption{The phase space for the dijet rate in the $E_{T1}-E_{T2}$ plane.}
    \label{fig:phase}
\end{figure}

To better understand the origin of the divergence in
$\sigma'(\Delta)$, we consider the value of the jet transverse energy
difference $E_{T1}-E_{T2}$. In the specific case of dijet
production in DIS, with an incoming quark $p$, at the Born level, a
dijet event consists only of two outgoing hard partons $p_1$ and
$p_2$, in this case a quark and a gluon. Their transverse momenta
$\vec p_{t1}$ and $\vec p_{t2}$ are back-to-back, so that, defining
$E_{Ti}=|\vec p_{ti}|$, we have $E_{T1}=E_{T2}$.  After emission of a
soft gluon $k$ not clustered with any of the two outgoing partons, we
obtain from transverse momentum conservation
$E_{T1}-E_{T2}\simeq|k_x|$, where $k_x$ is the component of the gluon
$\vec k_t$ parallel to $\vec p_{t1}$.  When considering
$\sigma'(\Delta)$, $E_{T1}$ is forced to lie on the line
$E_{T1}=E_m+\Delta$, but from kinematics $E_{T1}=E_{t2}+|k_x|$.
These two lines should then intersect somewhere in the allowed phase
space $E_{T1}>E_{T2}>E_m$, and this is possible only for
$|k_x|<\Delta$, as can be seen from fig.~\ref{fig:phase}.
Squeezing soft radiation is known to give rise to large logarithms in
fixed order calculations. Indeed, if we consider the emission of a
soft gluon collinear to the incoming quark together with the
corresponding virtual correction, we obtain
\begin{equation}
  \label{eq:sigma-NLO}
  \sigma'(\Delta)=\sigma'_0(\Delta)\times
\left(1-\frac{\alpha_s C_F}{\pi}\ln^2\frac Q\Delta\right)\>,
\end{equation}
so that the slope $\sigma'(\Delta)$ diverges for $\Delta\to 0$. A
quick solution of the problem can be achieved by choosing kinematical
cuts such that the Born slope $\sigma'_0(\Delta)\to 0$ for $\Delta\to
0$. This can be achieved for instance by imposing asymmetric $E_{T}$
cuts\cite{H1}, or choosing a suitable range for the dijet invariant
mass and rapidities\cite{ZEUS-multijet}.  However, the actual solution
relies in the all-order resummation of soft gluon effects, as will be
explained in the next section.

\section{Resummation}
\label{sec:resum}
Considering an arbitrary number of partons one finds $
E_{T1}-E_{T2}\simeq|\sum^\prime_i k_{ix}|$, where the primed sum runs
on all partons not clustered with the jets. Introducing $S(\vec k_t)$,
the probability that $\sum'_i \vec k_{ti}=\vec k_t$, we have
$\sigma'(\Delta)=\sigma'_0(\Delta) W(\Delta)$, where the K-factor
$W(\Delta)$ is in general given by
\begin{equation}
  \label{eq:W}
  W(\Delta)=\int d^2\vec k_t \>S(\vec k_t)\> \Theta(\Delta-|k_x|)=
  \frac 2\pi \int\frac{db}{b}\sin(b\Delta)\Sigma(b)\>.
\end{equation}
For emissions soft and collinear to the incoming quark, no secondary
partons are clustered with $p_1$ or $p_2$, so that an all-order
resummation of such contributions yields
$\Sigma(b)=\exp[-C_F\frac{\alpha_s}{\pi}\ln^2 b]$. From
eq.~(\ref{eq:W}) we have then $W(\Delta)\sim \Delta$ for small
$\Delta$, which roughly corresponds to the behaviour seen in the data.

In general, $\Sigma(b)=\exp[Lg_1(\alpha_s L)+g_2(\alpha_s L)+\alpha_s
g_3(\alpha_s L)\ldots]$, with $L=\ln b$, and we aim at next-to-leading
logarithmic (NLL) accuracy, i.e. the knowledge of $g_1$ and $g_2$.
In order to perform an all-order resummation, one need to reorganise
the PT series and collect all contributions to $\Sigma(b)$ up to a
given accuracy. First of all one can see that at NLL accuracy,
$\Sigma(b)$ can be interpreted as the probability that $|\sum^\prime_i
k_{ix}|<e^{-\gamma_E}/b$. Furthermore, given a variable $V$, a generic
function of all final state momenta, multi-parton matrix elements and
phase space constraints can be approximated by using the following
general properties:
\begin{enumerate}
\item
  \label{it:SC}
  all leading logarithms (LL) originate from configurations of
  soft and collinear gluons for which the hardest emission $k_1$
  dominates, that is $V(k_1,\dots,k_n)\simeq V(k_1)$;

\item 
  \label{it:HC}
  hard-collinear and soft-large-angle emissions give a NLL
  contribution, and again one can approximate $V(k_1,\dots,k_n)\simeq
  V(k_1)$;
  
\item 
  \label{it:multi}
  the fact that $V(k_1,\dots,k_n)\neq V(k_1)$ needs only to be taken
  into account for soft and collinear emissions; these ``multiple
  emission effects'' are NLL, and can be treated either analytically
  with a suitable integral transform, or numerically with Monte Carlo
  (MC) techniques;

\item 
  \label{it:ind}
  secondary splittings are accounted for by taking the QCD coupling in
  the physical CMW scheme, at a scale of the order of the transverse
  momentum of the parent parton, i.e. $\alpha_s\to \alpha_s(k_t)$.
\end{enumerate}
The strongest implication of these four statements is that at NLL
accuracy soft and/or collinear (SC) partons can be considered as
emitted independently from the hard parton antenna, in spite of the
fact that QCD is a non-abelian gauge theory. This crucial
simplification is known as the ``independent emission'' approximation,
and leads to exponentiation of leading logarithms, and factorisation
of NL logarithms. If we assume the validity of independent emission,
we obtain
\begin{equation}
  \label{eq:sigma-global}
  \Sigma(b)=\frac{f(1/b)}{f(Q)}\otimes e^{-R(b)}\>,
  \qquad R(b)=\int [dk] \>w(k)\> \Theta\left(|k_x|-e^{-\gamma_E}/b\right)\>, 
\end{equation}
where the 'radiator' $R(b)$ is (minus) the contribution to $\Sigma(b)$
from a single SC gluon $k$ emitted with probability $w(k)$. The ratio
of parton densities $f(1/b)/f(Q)$ is a general feature of resummations
involving incoming partons, and is due to the fact that the scale of
the incoming parton density is set by the upper limit of observed
parton transverse momenta, which in this case is of order $1/b$.

Unfortunately, independent emission approximation does not hold for
all variables, but only for those who are recursively infrared and
collinear (rIRC) safe and (continuously) global\cite{CAESAR}. The lack
of either of these two properties causes one (or more) of the above
statements not to be true.

For instance, for the three-jet rate in the JADE algorithm, which is
global but not rIRC safe, statement~\ref{it:SC} is false, since there are LL
contributions originated by multiple emissions, with the consequence
that leading logarithms do not exponentiate\cite{JADE}.

Globalness means simply that $V$ is sensitive to emissions in the
whole of the phase space. This is not the case for $E_{T1}-E_{T2}$,
since only partons that are not clustered with the two hard jets
contribute to the dijet transverse energy difference.  The first
consequence of non-globalness is that secondary splittings cannot be
accounted for by simply setting the proper scale for $\alpha_s$,
because new NLL contributions, called ``non-global
logs''\cite{Dassal1}, arise when a cascade of energy ordered partons
outside the measure region emits a softer gluon inside.  In this case
phase space boundaries have to be taken into account exactly for all
emissions, so that one has to rely on MC methods.  Moreover, due to the
complicated colour structure of multi-gluon matrix elements,
non-global logs at the moment can be resummed only in the large $N_c$
limit.

In our case, if particles are clustered into jets with a cone
algorithm, after having specified a procedure to deal with overlapping
cones, one has that to NLL accuracy a jet consists of all particles
that flow into a cone around $p_1$ or $p_2$, and
equation~(\ref{eq:sigma-global}) gets modified as follows\cite{BDetdif}:
\begin{equation}
  \label{eq:sigma-nonglobal}
  \Sigma(b) = \frac{f(1/b)}{f(Q)}\otimes e^{-R(b)} \times S[t(b)]\>,
  \qquad t(b)=\frac{1}{2\pi}\int_{1/b}^{Q}\frac{dk_t}{k_t}\alpha_s(k_t)\>,
\end{equation}
where $S(t)$ represents the contribution of non-global logs, and can be
computed with a MC procedure as a function of the evolution variable $t$.

If jets are clustered with the $k_t$ algorithm\cite{ktpaper}, further
complications arise. In this case the fact that a particle belongs to
a jet depends on all other emitted particles.  This has interesting
implications for non-global logs. Dasgupta and Salam\cite{Dassal2}
have shown that the dominant contribution to non-global logs arises
when the observed gluon is close to the boundary of the measure
region.  Appleby and Seymour\cite{AS1} noted then that the $k_t$ algorithm
requirement forces such gluons to be clustered with gluons outside the
measure region, thus reducing the magnitude of non-global logs.

One can also ask whether for the independent emission contribution to
$\Sigma(b)$, the exponential form of equation~(\ref{eq:sigma-global})
is still valid. This is true for cones (the ``unclustered'' case)
since the phase space constraint is factorised, but should be checked
for the $k_t$ algorithm (the ``clustered'' case). In order to answer
this question we considered a much simpler variable, the transverse
energy flow away from the jets.  Given a pair of hard jets (taken
back-to-back for simplicity), one defines a region $\Omega$ away from
the jet axis (which in $e^+e^-$ annihilation roughly coincides with the thrust
axis), and the away-from-jet $E_T$ flow
\begin{equation}
  \label{eq:etflow}
  E_{T,\Omega}=\sum_{i\in \Omega} k_{ti}\>,
\end{equation}
with $k_{ti}$ the transverse momentum of the $i$-th {\it jet} with
respect to the jet axis\cite{Dassal2,AS1}. The quantity
$\Sigma(Q,Q_\Omega)$, the probability that $E_{T,\Omega}<Q_\Omega$, in
the region $Q_\Omega \ll Q$ is sensitive only to soft gluons at large
angles. One then wishes to resum LL contributions to
$\Sigma(Q,Q_\Omega)$, whose order is $\alpha_s^n \ln^n(Q/Q_\Omega)$.

At LL accuracy $\Sigma(Q,Q_\Omega)$ is given by
\begin{equation}
  \label{eq:Sigma-Om}
  \Sigma(Q,Q_\Omega) = \Sigma_{\Omega,P}(t)\cdot  S(t)\>,
\end{equation}
where $t\equiv t(Q_\Omega)$ is the evolution variable defined in
equation~(\ref{eq:sigma-nonglobal}). Gluons emitted directly from the
two hard partons give rise to $\Sigma_{\Omega,P}(t)$, while non-global
logs are embodied in $S(t)$.  Appleby and Seymour\cite{AS1} assumed
that also for the clustered case $\Sigma_{\Omega,P}(t)=e^{-R(t)}$,
that is the single gluon contribution to $\Sigma(Q,Q_\Omega)$
exponentiates.  This naive expectation can be motivated by the fact
that if emissions are assumed to be independent, multiple emission
effects are usually relevant only for soft and collinear gluons
(statement~\ref{it:multi}). However, since $E_{T,\Omega}$ is
manifestly non-global, one cannot exclude multiple emission effects
coming from soft gluons at large angles.  Consider for instance two
gluons $k_1$ and $k_2$, with $\omega_1 \gg \omega_2$, $k_2$ inside
$\Omega$ and $k_1$ outside. It is possible that the jet algorithm
cluster $k_2$ with $k_1$, thus spoiling the exponentiation of the single
gluon result\cite{ktletter}.  This is indeed seen when comparing the
resummed expression for the differential distribution $\sigma^{-1}
d\sigma/dL$, with $L=\ln (Q_\Omega/Q)$, with the NLO program
EVENT2\cite{DISENT}.
\begin{figure}[t]
  \centering
    \centerline{\epsfxsize=.7\textwidth\epsfbox{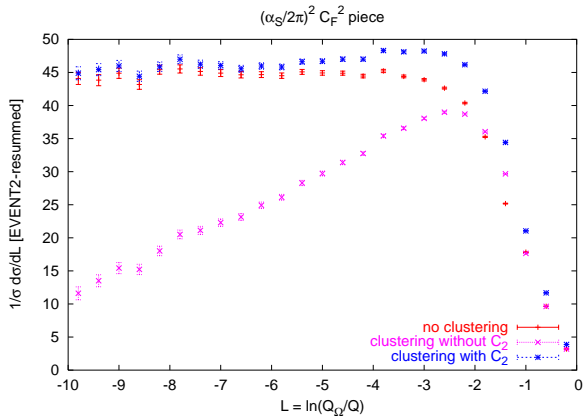}}
    \caption{The coefficient of $(\alpha_s/2\pi)^2 C_F^2$ for the difference
      between resummed and NLO predictions for
      $\sigma^{-1}d\sigma/dL$, as explained in the text. Jets are
      clustered with the $k_t$ algorithm with $R=1$, and $\Omega$
      is a rapidity slice with $\Delta \eta =1$. $C_2$ is the
      correction that needs to be applied at order $\alpha_s^2$ to
      take into account multiple emission effects.}
  \label{fig:nlo-resum}
\end{figure}
 If LL are correct, the difference of the two
distributions should go to a constant for large (negative) $L$. This
happens for the coefficient of $C_F C_A \alpha^2_s$, indicating that
both in the unclustered and the clustered case non-global
logarithms are correctly taken into account.  If one however uses for
the clustered case the same expression for $\Sigma_{\Omega,P}$ as for
the unclustered case, one finds a discrepancy in the coefficient of
$C_F^2 \alpha^2_s$, the plot of fig.~\ref{fig:nlo-resum}. In the
clustered case $\Sigma_{\Omega,P}$ must be corrected by taking into
account the fact that the softest gluon, in spite of the fact that is
emitted in $\Omega$, can be nevertheless clustered with the hard jets,
and therefore does not contribute to $\Sigma_{\Omega,P}$. Even for
primary emissions this constitutes a LL contribution.
The resummation of these multiple emission effects can be performed
with the same MC used for non-global logs, and is the last bit that is
needed to resum the dijet rate  with the $k_t$
algorithm at NLL accuracy.

\section{Phenomenology of dijet observables}
\label{sec:dijet}
Instead of going through the details of the resummed calculation for
the dijet rate in DIS, we discuss what information on QCD dynamics we
can gain from allowing symmetric $E_T$ cuts. Since in the region
$\Delta \to 0$ we are almost in a three-jet configuration, this
measurement is complementary to three-jet event shapes, and allows one
to investigate the coherence properties of QCD emission from a three
parton antenna.  This observable has also the advantage that NP
corrections are smaller than in event shapes distributions. This is
mainly due to the fact that setting $\Delta\to 0$ does not put a
direct veto on emitted parton transverse momenta, but rather on their
vector sum. We have then emissions with large transverse momenta
contributing also in the extreme $\Delta\to 0$ region. This fact also
affects the magnitude of non-global logs, which here\cite{BDetdif}
give a correction of order 10\%, while for event shapes\cite{DSreview}
their contribution can be as large as 30\%.

One can consider other observables that have the same resummation
as the dijet rate and can be more easily handled theoretically or
experimentally. Among these are the distributions in the transverse
energy difference $\Delta=E_{T1}-E_{T2}$ and in the azimuthal angle
$\Delta\phi$ between the jets. For any of these variables $V$, one studies
\begin{equation}
  \label{eq:sigmadif}
  \frac{1}{\sigma}\frac{d\sigma}{dV}=\frac{d}{dV}\Sigma(V)\>.
\end{equation}
As already stated, no further effort is needed to resum these
distribution, since $\Sigma(V)=\sigma_0(0)W(V)$, where $W$ is the same as in
eq.~(\ref{eq:W}). Since from the previous analysis we have seen
that $\Sigma(V)\sim V$ for small $V$, one expects any of these
differential distributions to approach a constant for $V\to 0$. This
is what is already seen in the data for the $\Delta\phi$ distribution
in hadronic dijet production at the Tevatron\cite{D0}. This observable is
particularly interesting because it gives access to the QCD {\it fifth
  form factor}, which is present when there are at least four hard 
emitting partons\cite{DMfifth}.  

\begin{figure}[t]
  \centering
    \centerline{\epsfxsize=.7\textwidth\epsfbox{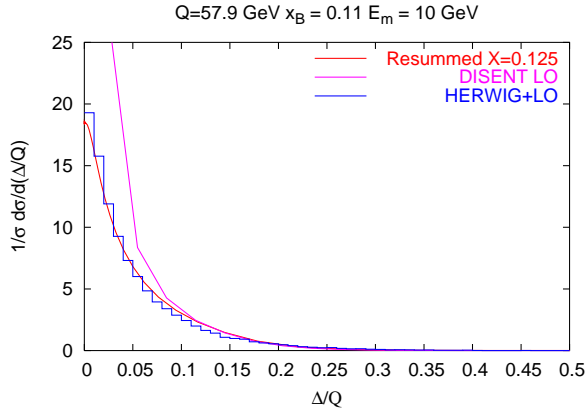}}
    \caption{Preliminary plots for the transverse energy differential
      distribution in DIS. The parameter $X$ is used to estimate NNLL
      effects.  }
\label{fig:herwig}
\end{figure}

A plateau for small $\Delta$ is also seen in the differential
distribution in the dijet transverse energy difference in DIS computed
with HERWIG\cite{HERWIG}. Fig.~\ref{fig:herwig} shows the comparison
between the resummed result and that obtained from HERWIG.  In this
case jets are identified with the $k_t$ algorithm, and in the
resummation non-global logs have been approximated with their value
for the cone algorithm, which has been taken as an upper bound. The
pretty good agreement between the two predictions, together with the fact
HERWIG does not fully contain the matrix elements giving rise to
non-global logs, suggests that the observable is dominated by soft and
collinear emissions.

Another interesting possibility could be that of making
$E_{T1}-E_{T2}$ global by modifying the particle recombination scheme.
Actually, in a massless $E_0$ scheme, one finds that emissions
clustered with the jets do contribute to $E_{T1}-E_{T2}$, so that one
can use the program CAESAR\cite{CAESAR} to obtain automatically the resummed
expression for its distribution.

The concluding remark of this overview is that pushing measurements in
phase space regions where fixed order predictions are not expected to
give an accurate description of the data opens up the possibility to
investigate properties of QCD otherwise unaccessible.  In particular,
cross sections for jets with almost equal $E_T$'s represent a yet
poorly explored field, which could be complementary to the traditional
event-shape measurements.

\section*{Acknowledgements}
This work would not have been possible without the collaboration of
Gennaro Corcella and Mrinal Dasgupta. I am grateful to the organisers
for the invitation and for the possibility to enjoy the beauty of
Ringberg castle and Tegernsee together with my wife and my baby
daughter.

\end{document}